\documentclass[aps,pra,epsfigure,twocolumn,longbibliography]{revtex4-1}
\usepackage{dcolumn}    
\usepackage{bm} 
\usepackage{graphicx}
\usepackage{amsmath}    
\usepackage{latexsym}
\usepackage{amsfonts}   
\usepackage{amssymb}
\usepackage{array}      
\usepackage{epsfig}
\usepackage{txfonts}
\usepackage{color}
\usepackage[colorlinks=true,linkcolor=blue,urlcolor=blue,citecolor=blue,pdfusetitle]{hyperref}
\usepackage{hyperref}
\usepackage[dvipsnames]{xcolor}

\newcommand{\more}{...more\\more\\more\\more\\more}
\newcommand{\ket}[1]{\left\vert#1\right\rangle}
\newcommand{\bra}[1]{\left\langle#1\right\vert}

\begin{document}
\title{Energetic cost of quantum control protocols}

\author{Obinna Abah,$^1$ Ricardo Puebla,$^1$ Anthony Kiely,$^{2,3}$ Gabriele De Chiara,$^1$ Mauro Paternostro,$^1$ and Steve Campbell$^{4,5}$}
\affiliation{
$^1$Centre for Theoretical Atomic, Molecular and Optical Physics, School of Mathematics and Physics, Queen's University Belfast, Belfast BT7 1NN, United Kingdom,\\
$^2$Departamento de Qu\'{\i}mica F\'{\i}sica, UPV/EHU, Apdo 644, 48080 Bilbao, Spain,\\
$^3$Department of Physics, University College Cork, Ireland\\
$^4$School of Physics, University College Dublin, Belfield Dublin 4, Ireland\\
$^5$School of Physics, Trinity College Dublin, Dublin 2, Ireland
}

\begin{abstract}
We quantitatively assess the energetic cost of several well-known control protocols that achieve a finite time adiabatic dynamics, namely counterdiabatic and local counterdiabatic driving, optimal control, and inverse engineering. By employing a cost measure based on the norm of the total driving Hamiltonian, we show that a hierarchy of costs emerges that is dependent on the protocol duration. As case studies we explore the Landau-Zener model, the quantum harmonic oscillator, and the Jaynes-Cummings model and establish that qualitatively similar results hold in all cases. For the analytically tractable Landau-Zener case, we further relate the effectiveness of a control protocol with the spectral features of the new driving Hamiltonians and show that in the case of counterdiabatic driving, it is possible to further minimize the cost by optimizing the ramp.
\end{abstract}
\date{\today}
\maketitle

\section{Introduction}
The inherent fragility of quantum systems necessitates that we develop methods to coherently control their evolution~\cite{STAreview, STAreview2}. The need for high precision control is evidently ubiquitous; the study of how and why peculiar quantum properties manifest requires techniques that allows for the careful manipulation of these systems. While a variety of techniques have been developed for many types of quantum system~\cite{STAreview, STAreview2, BasonNatPhys2012, Zhang2013PRL}, often neglected has been the associated resources needed to achieve this high degree of control. While such an omission is evidently justified when one is solely interested in studying a particular quantum phenomenon, it is vital to account for such expenditures when developing novel technologies that exploit these quantum features. Indeed, recently the application of control techniques that can achieve an effective adiabatic dynamics in a finite time, called ``shortcuts-to-adiabaticity"~\cite{STAreview, STAreview2}, has been shown to be highly effective in a diverse range of settings including quantum gates~\cite{Santos2015SciRep}, quantum games~\cite{ShersonNature, SelsPRA2018} nano-scale thermodynamic cycles~\cite{GooldSciRep, Abah2017EPL, Abah2018PRE, BarisPRE2019, LiNJP2018}, open quantum systems~\cite{VacantiNJP, PolettiPRA2016, KosloffPRL2019, AdolfoArXiv}, manipulating critical many-body systems~\cite{delCampoPRL2012, CampbellPRL2015}, and quantum precision measurements~\cite{PangNatComms}. This further highlights the importance of understanding the additional resources required to achieve precise control in a quantitative manner.

The question of how to quantify the necessary resources to control a quantum system using a particular protocol has recently become a topic of intense research activity (indeed, in the context of thermodynamic cycles the issue becomes more subtle since any additional energy which is not dissipated can in principle be recycled and act as a catalyst~\cite{Kosloff2017}). The variety of ways in which a particular set-up can be coherently controlled has led to a plethora of definitions~\cite{Muga2017PRA, MugaNJP2018, HorowitzPRL, BarisPRE2019, CalzettaPRA, Funo2017PRL, Chen2016JPCA, Demirplak2008, Zheng2016PRA, Campbell2017PRL, Santos2015SciRep, Santos2017, HerreraPRA, Chen2010PRA, Abah2017EPL, ImpensSciRep, GooldSciRep, Abah2018PRE, Abah2019PRE, Bravetti2017PRE, delCampoArXiv, MortensenNJP, SelsPRX}. Nevertheless, since many of these quantifiers invariably share some common traits, it leads to a natural question: Which control protocols are the most resource intensive?

In this work we begin to tackle this issue by employing the cost measure introduced in Ref.~\cite{Zheng2016PRA} and, through it, quantitatively and qualitatively compare and contrast several different coherent control protocols. For a fixed protocol duration, $\tau$, naturally, one must choose a figure of merit with which to judge the success of the process. Here we choose the target state fidelity, $\mathcal{F}\!=\!\vert\bra{\psi(\tau)} \Psi\rangle\vert^2 \!\to\! 1$, where $\ket{\psi(\tau)}$ is the evolved state of our system using a particular control protocol and $\ket{\Psi}$ is the target state we are aiming to achieve. By fixing the quantifier of cost and examining the paradigmatic settings of the Landau-Zener model, which serves to elucidate the control needs of critical many-body systems~\cite{delCampoPRL2012}, the parametric quantum harmonic oscillator, and the Jaynes-Cummings model, we show that a consistent hierarchy of costs can emerge. We find that techniques that suppress all non-adiabatic excitations are generally energetically costly protocols and we relate this to the effect that these more resource intensive techniques have on the energy spectrum of the controlled system. However, we show that the cost can be minimized by exploiting the freedom in choosing how one ramps the system. Furthermore, we establish that optimal control and inverse engineering protocols are generally less energetically costly.

The manuscript is organized as follows. In Sec.~\ref{prelim}, we outline the basic tools utilized throughout this work. Sec.~\ref{results} quantitatively analyzes the energetic cost of control for three paradigmatic settings: ramping the ground state of the Landau-Zener model through its avoided crossing, compressing the thermal state of a quantum harmonic oscillator, and tuning the light-matter interaction strength in the Jaynes-Cummings model. Finally, in Sec.~\ref{conclusion} we draw our conclusions and provide some further discussions.

\section{Preliminaries}
\label{prelim}
Controlling quantum systems such that an effective adiabatic dynamics is realized in a finite time can be achieved through a variety of techniques~\cite{STAreview, STAreview2, DeffnerReview}. In this work, we will focus on several of the most prevalent of such protocols for a given situation and, for brevity, we refer to the comprehensive reviews on the topics for a detailed discussion of their derivations and implementations~\cite{STAreview, STAreview2}. Counter-diabatic (CD) or transitionless quantum driving is one such method that involves adding an additional correction term to the bare Hamiltonian, $H_0$, such that the resulting dynamics exactly tracks the corresponding adiabatic dynamics~\cite{Demirplak2003, Demirplak2005, Berry2009JPA}. If one is only interested in controlling populations, then with a suitable choice of phase~\cite{Santos2017} this can be achieved through the CD term
\begin{equation}
\label{berryeq}
H_\text{CD}=i \sum_n \ket{\partial_t \psi_n(t) }\bra{\psi_n(t)},
\end{equation}
where $\ket{\psi_n(t)}$ are the eigenstates of the bare system Hamiltonian one is interested in manipulating and where we assume units such that $\hbar\!=\!1$. An oft-cited drawback of this approach is that the resulting correction term can be highly non-local~\cite{Muga2010JPB, delCampoPRL2012, Campbell2017PRL} and therefore difficult to implement. However, for certain systems by exploiting a unitary transformation, the total $H_0+H_\text{CD}$ Hamiltonian can be re-expressed in the so-called local counter-diabatic (LCD) form, $H_\text{LCD}$, where perfect final target state fidelity is still achieved~\cite{delCampo2013PRL}. Crucially, though, $H_\text{LCD}$ does not involve any complex non-local operators and is instead constructed using the same operators that appear in $H_0$. Another drawback of the CD approach is that, in principle, it requires full spectral knowledge. Thus often for complex systems where complete spectral information is not available, alternative approaches must be employed. In this work, when possible, we will also consider other more heuristic  protocols, optimal control theory (OC)~\cite{CRAB1, CRAB2, CRAB3} and inverse engineering (IE)~\cite{STAreview}, and compare the resource intensiveness of their implementation.

Our aim is to both qualitatively and quantitatively assess the cost of implementing these protocols, which is a topic that has ignited significant interest recently~\cite{Muga2017PRA, MugaNJP2018, HorowitzPRL, BarisPRE2019, CalzettaPRA, Funo2017PRL, Chen2016JPCA, Demirplak2008, Zheng2016PRA, Campbell2017PRL, Santos2015SciRep, Santos2017, HerreraPRA, Chen2010PRA, Abah2017EPL, ImpensSciRep, GooldSciRep, Abah2018PRE, Abah2019PRE, Bravetti2017PRE, delCampoArXiv, MortensenNJP, SelsPRX}. Indeed as discussed in Ref.~\cite{STAreview2} the notion of the cost has been somewhat loosely employed and therefore different quantifiers probe different aspects of the system's energy or its interactions. In this regard, we are in principle free to choose or define any meaningful quantifier we wish. However, we must ensure that whichever approach we use provides a sound basis for drawing a comparison. For example, simply determining the average energy of the state $\bra{\psi(t)} H_0 \ket{\psi(t)}$ is insufficient as the CD approach will appear to be free as the instantaneous energy will be identical to the adiabatic energy. Here we mainly focus on the cost as defined by Zheng {\it et al}~\cite{Zheng2016PRA} and use the norm of the Hamiltonian to define the instantaneous cost of the evolution
\begin{equation}
\label{cost}
\partial_t \mathcal{C} = \| H_k \|
\end{equation}
using the Frobenius norm, where $H_k$ is the total Hamiltonian used in determining the evolution and $k$ is used to distinguish the various protocols. Notice that Ref.~\cite{Zheng2016PRA} was concerned with determining the additional resources necessary to implement CD only and therefore defined the cost in terms of the additional energy added to the bare Hamiltonian. Here the cost is related to the norm of the full Hamiltonian implemented during the evolution and thus accounts for the {\it total} energy of the process, rather than only defining the energetic cost to achieve the control protocol.

A few important remarks are in order. Firstly, as we shall focus on unitary dynamics, any additional energy resources employed are not dissipated. This is a subtle issue that is particularly relevant if one wishes to extend our analysis to the performance of thermodynamic cycles~\cite{GooldSciRep, Abah2017EPL, Abah2018PRE, LiNJP2018, BarisPRE2019}, as it is possible for the additional energy requirements invested in achieving coherent control to be recycled, see e.g. Refs.~\cite{Muga2017PRA, Kosloff2017}.  Secondly, to ensure a fair comparison, we explicitly account for the bare Hamiltonian contribution to the energy requirements, thus ensuring that no evolution is free. By employing Eq.~\eqref{cost} our analysis essentially focuses on the intensity of all the driving fields in achieving high fidelity control, and while we expect the qualitative behavior to persist for other definitions of cost, this question requires nevertheless a systematic study in itself.

\section{Case Studies}
\label{results}
\subsection{Landau-Zener Model}
To begin, we consider a single spin in a time-dependent field
\begin{equation}
\label{LZham}
H_0= \Delta \frac{\sigma_x}{2} + g(t) \frac{\sigma_z}{2}.
\end{equation}
In what follows we will assume that the system is initialized in its ground state with $g(0)\!=\!-0.2$ and we wish to evolve through the avoided crossing to $g(\tau)\!=\!0.2$. Our goal is to estimate the energy used to achieve this evolution under the condition that the fidelity at the start and end of the process is close to unity. Depending on the control protocol employed we may allow for the transient to leave the ground state manifold. 

For OC the fastest approach is given by a bang-off-bang (BOB) pulse~\cite{Caneva2009PRL, Hegerfeldt2013PRL, PoggiEPL}, where the system is suddenly and strongly quenched, followed by a free evolution with no field, and finally a reverse sudden quench is applied
\begin{equation}
g_\text{BOB}(t) = \begin{cases}
~ g_Q,\qquad &t=0, \\
~ 0, \qquad &0<t<\tau, \\
 -g_Q, \qquad &t=\tau,
\end{cases}
\label{eq:bangoffbang}
\end{equation}
with $g_Q \!\gg\! g(0)$ (in our simulations $g_Q\!=\!100$ is sufficient). This approach is effective when the evolution time is given by the quantum speed limit (QSL) time, $\tau_\text{QSL}$~\cite{DeffnerReview, Frey2016, Caneva2009PRL, Hegerfeldt2013PRL}. However we will also consider more general approaches valid for $\tau\!>\!\tau_\text{QSL}$ later. By focusing on initial and target ground states of the Landau-Zener model, Eq.~\eqref{LZham}, the QSL time can be found fully analytically and is given by~\cite{Hegerfeldt2013PRL} 
\begin{equation}
\cos \left(\frac{\Delta}{2} \tau_\text{QSL}\right) = \vert  \alpha_i \alpha_t  \vert + \vert  \beta_i \beta_t  \vert,
\end{equation}
where $\alpha_{i(t)}$ and $\beta_{i(t)}$ correspond to the $\sigma_z$ basis coefficients of the normalized initial (target) state, respectively.

\begin{figure*}[t]
{\bf (a)} \hskip0.65\columnwidth {\bf (b)} \hskip0.65\columnwidth {\bf (c)}\\
\includegraphics[width=0.65\columnwidth]{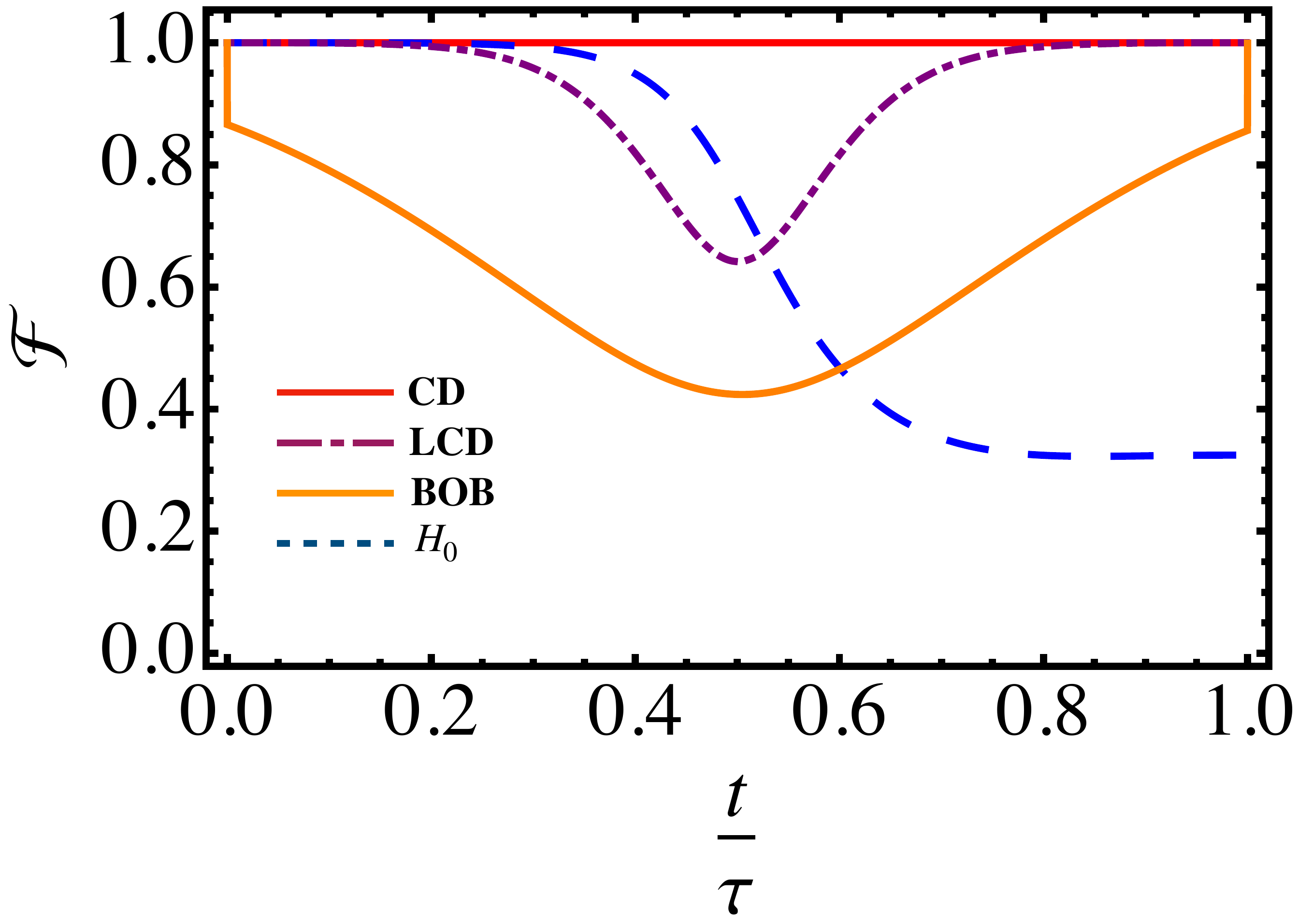}
\includegraphics[width=0.65\columnwidth]{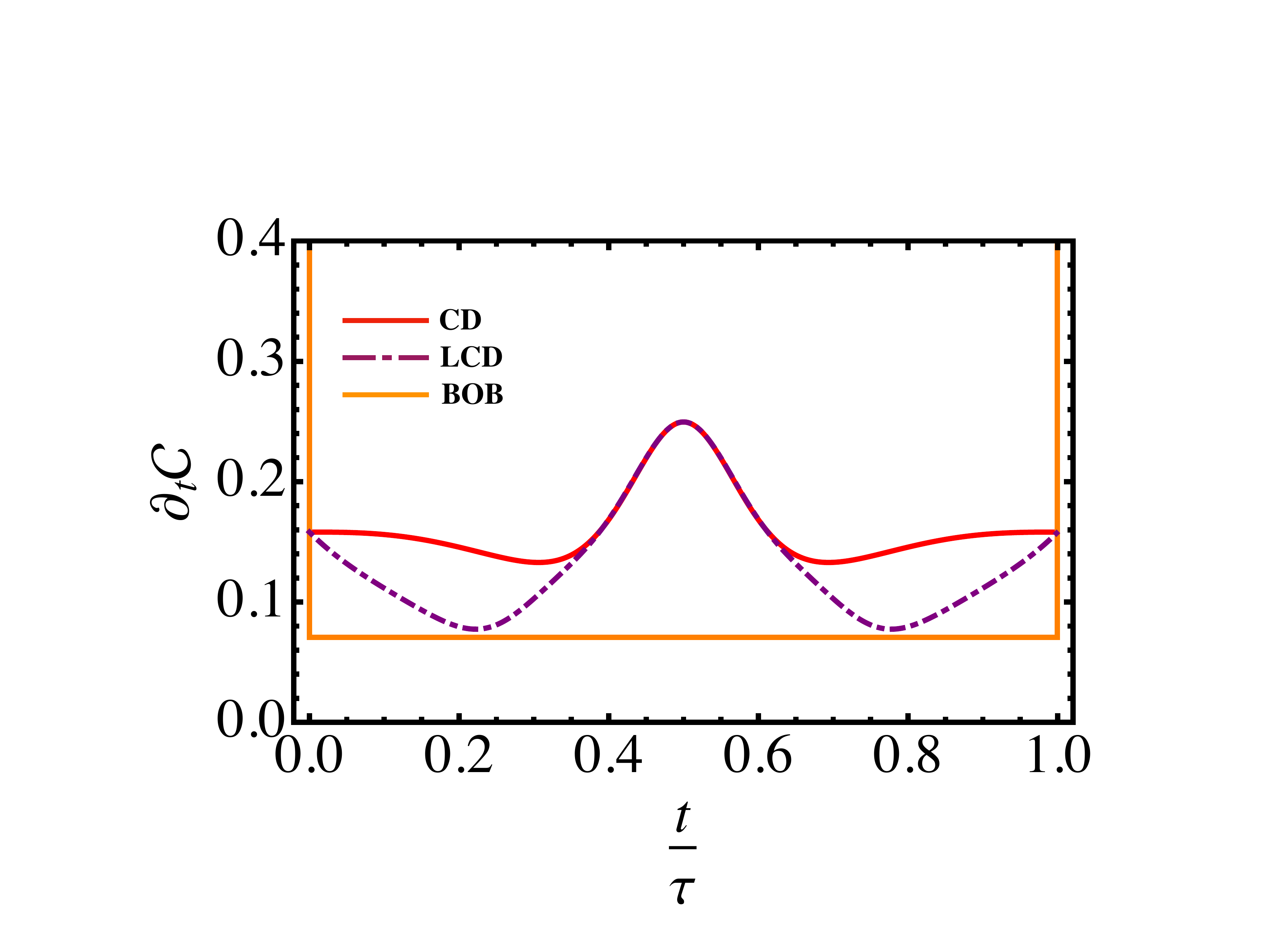}
\includegraphics[width=0.7\columnwidth]{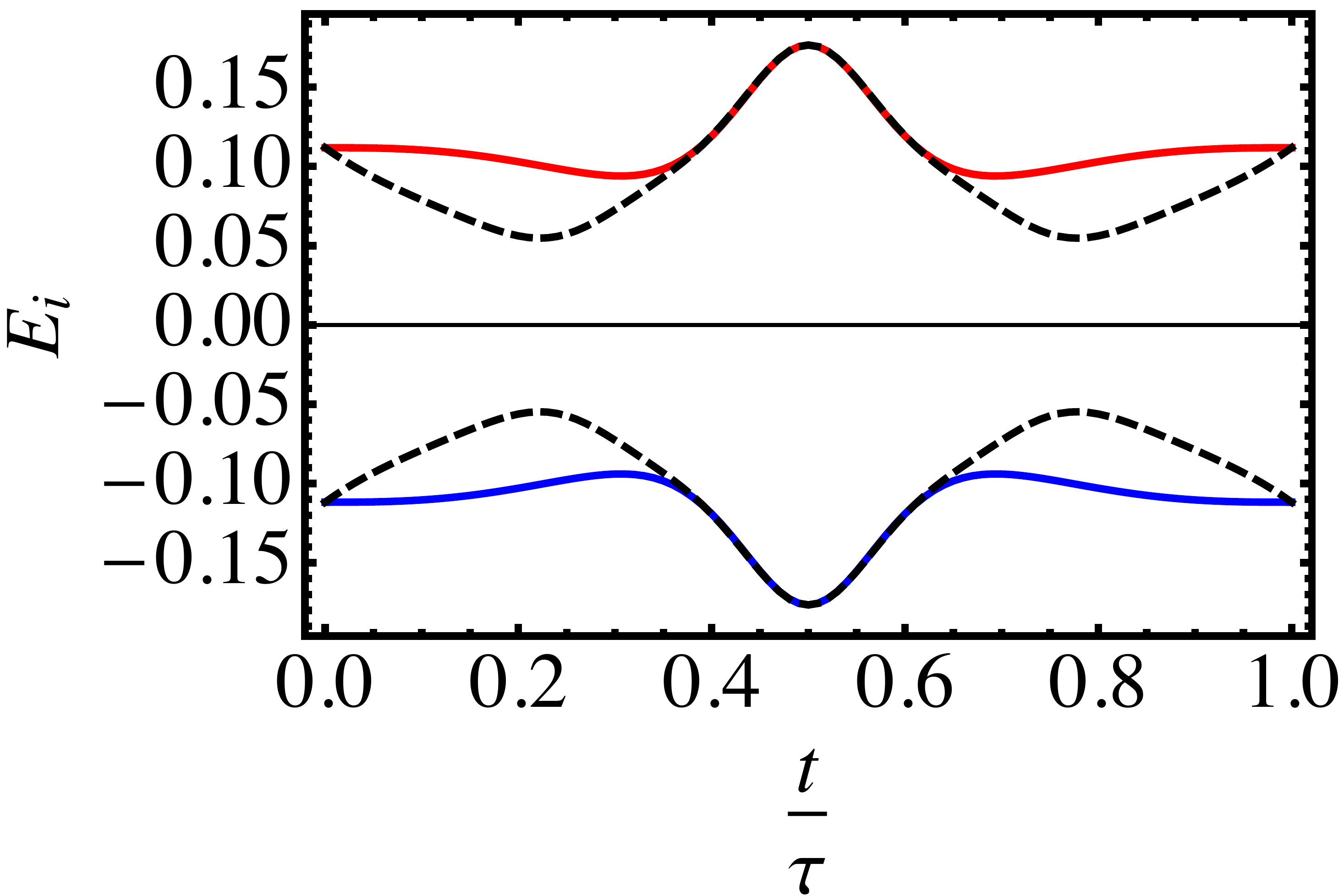}\\
{\bf (d)} \hskip0.65\columnwidth {\bf (e)} \hskip0.65\columnwidth {\bf (f)}\\
\includegraphics[width=0.65\columnwidth]{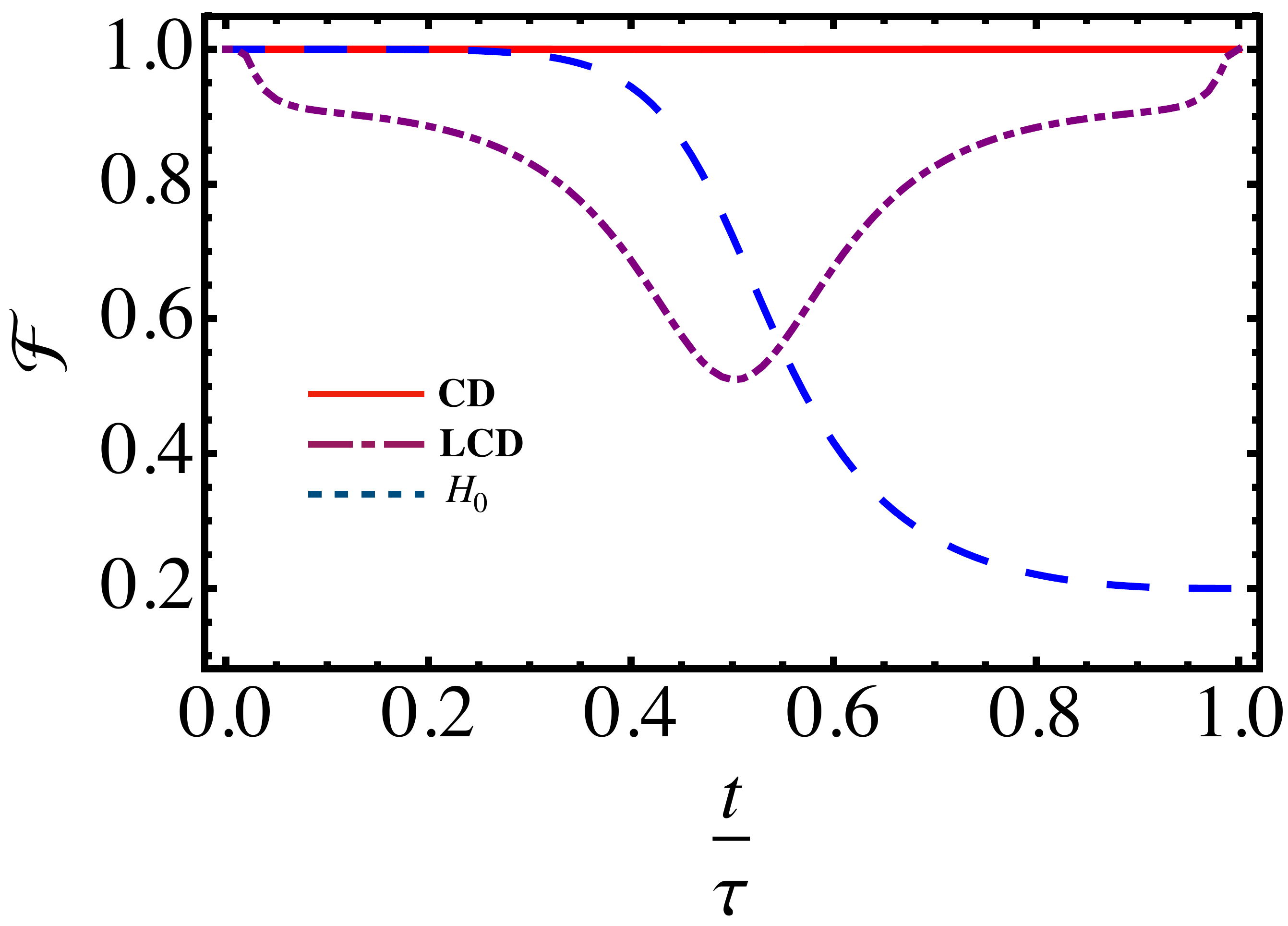}
\includegraphics[width=0.65\columnwidth]{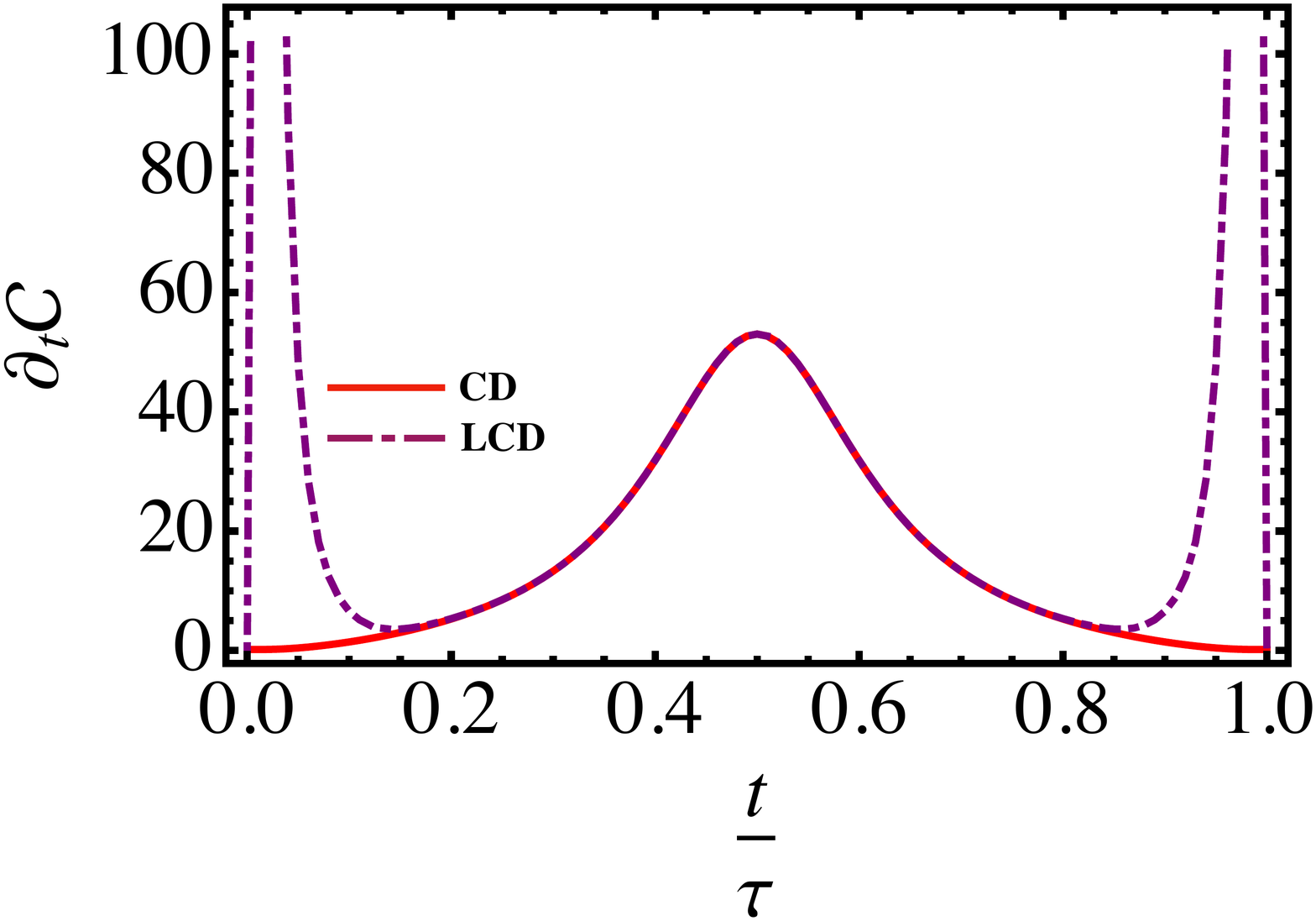}
\includegraphics[width=0.65\columnwidth]{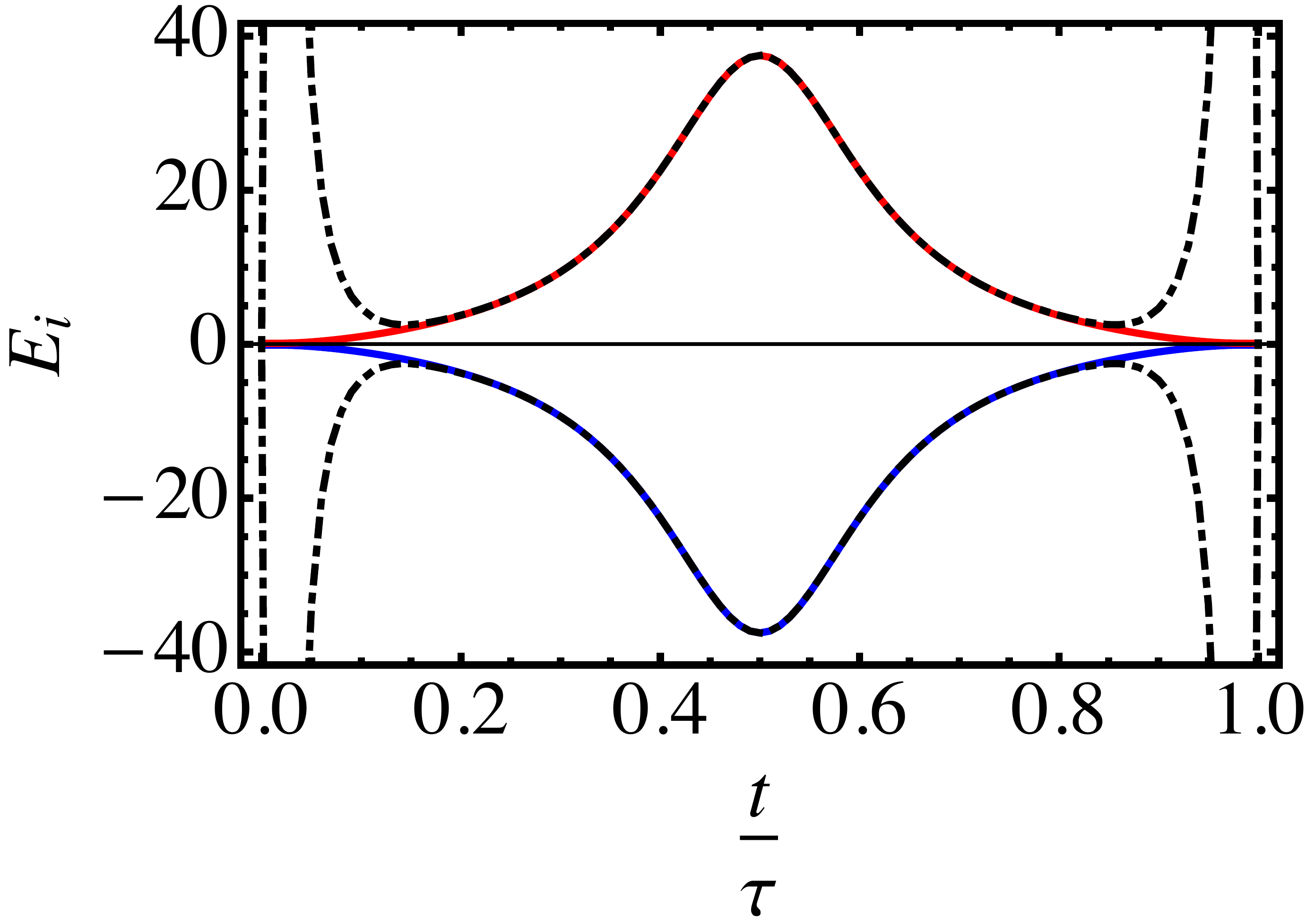}
\caption{{\bf (a)} Fidelity of various control protocols with the instantaneous ground state, assuming $g(t)$ takes the form in Eq.~\eqref{LCDramp}. Here we fix $\tau\!=\!\tau_\text{QSL}\!\approx\!22.14$. {\bf (b)} Corresponding instantaneous cost Eq.~\eqref{cost} for BOB (bottom-most, orange), CD (solid, red) and LCD (dot-dashed, purple). {\bf (c)} Energy spectra for the full CD Hamiltonian, $H_0+H_\text{CD}$, [solid colors] and the LCD Hamiltonian, $H_\text{LCD}$, [dashed, black]. {\bf (d)}-{\bf (f)} As for previous panels with $\tau\!=\!0.1$. In {\bf (e)} and {\bf (f)} we have truncated the vertical axis for clarity. In all figures we fix the energy gap $\Delta\!=\!0.1$ in Eq.~\eqref{LZham}}
\label{fig1}
\end{figure*}

The CD control field which must be added to the bare Hamiltonian is given by~\cite{Berry2009JPA}
\begin{equation}
\label{qubitTQD}
H_\text{CD} = -\frac{g'(t) \Delta}{2\left[\Delta^2 + g^2(t) \right]} \sigma_y.
\end{equation}
This control technique ensures that not only will the system be in the required state at the end of the protocol but it will also remain in the instantaneous eigenstate of the original Hamiltonian throughout. There is complete freedom in both the form of the ramp and its duration.

Turning to LCD, perfect target state fidelity can be achieved by making a unitary transformation of $H_0+H_\text{CD}$ to arrive at~\cite{PolettiPRA2016, Stefanatos2019PRA}
\begin{equation}
\label{qubitLCD}
H_\text{LCD} = P(t) \frac{\sigma_x}{2} + \left[g(t) -\dot{\eta}(t) \right] \frac{\sigma_z}{2},
\end{equation}
with $P(t)\!\!=\!\!\sqrt{\Delta^2+\dot{\theta}^2}$, $\theta\!\!=\!\!\text{arccot}\left[g(t)/\Delta\right]$, and $\eta(t)\!\!=\!\!\arctan(\dot{\theta}/\Delta)$. Notice that, as with OC, the shortcut is now achieved using a Hamiltonian that is of the same general form as the bare Hamiltonian, $H_0$. Unlike CD, where the form of the ramp can be completely arbitrary, for the LCD term to be effective a particular form of ramp is required, one with smooth start and end points, given by~\cite{PolettiPRA2016}
\begin{equation}
\label{LCDramp}
g(t) = g_0 + 10 g_d \left(\frac{t}{\tau}\right)^3 - 15 g_d \left(\frac{t}{\tau}\right)^4 + 6 g_d \left(\frac{t}{\tau}\right)^5.
\end{equation}
with $g_0\!=\!-0.2$ and $g_d\!=\!0.4$. Despite this, we can choose $\tau$ to be arbitrarily small and, in particular, smaller than the QSL time. This shows a key difference between OC, where an optimized path for varying $g(t)$ is found, and the LCD approach. On the one hand, we see that OC is bounded by the QSL when only the field is varied. On the other hand, with LCD we are also time-dependently varying the energy splitting, $\Delta$, via $P(t)$ in Eq.~\eqref{qubitLCD} and therefore we can drive the system faster than the QSL. It is important to notice that we `beat' the speed limit using CD and LCD because we are significantly altering the spectrum of the system. In essence, the more energy available to be imparted to the system the faster the evolution can be performed~\cite{Campbell2017PRL}. 

As mentioned, we will initialize our system in the ground state for $g(0)\!=\!-0.2$, the target state will be the ground state at $g(\tau)\!=\!0.2$ and we initially fix the gap $\Delta\!=\!0.1$. We will consider the BOB pulse performed at the quantum speed limit, $\tau\!=\!\tau_\text{QSL}\!\approx\! 22.14$, while for CD and LCD, as there are no constraints on how fast the protocol can be achieved we will consider, $\tau\!=\!\tau_\text{QSL}$ so as to compare faithfully with BOB, and $\tau\!=\!0.1$, i.e. extremely fast driving. For both CD and LCD we will employ the smooth ramp as given by Eq.~\eqref{LCDramp}.

In Fig.~\ref{fig1}(a) we examine the instantaneous fidelity, $\mathcal{F}\!=\!\vert\bra{\psi(t)} \phi \rangle\vert^2$, between the states $\ket{\psi(t)}$ evolved according to BOB, CD, LCD, and the bare Hamiltonian, with the corresponding instantaneous adiabatic state $\ket{\phi}$ of $H_0$ using the ramp Eq.~\eqref{LCDramp} with $\tau\!=\!\tau_\text{QSL}$. Clearly we see a qualitative similarity in the behavior of BOB and LCD, both protocols achieve the target state by evolving through partially excited states of the system. Fig.~\ref{fig1}{(b)} shows the corresponding instantaneous cost, Eq.~\eqref{cost}. Immediately, and somewhat expectantly as they involve strong additional control fields, CD and LQD are resource intensive approaches. Interestingly, the BOB protocol is by far the most efficient. With the exception of two strong pulses when driving at the QSL, the system consumes comparatively little energy. It is worth noting the dichotomy between the behavior of the cost for CD and LCD compared with BOB: the former ones are maximized at the avoided crossing while the latter is minimized. For CD, as discussed in Ref.~\cite{Campbell2017PRL}, the speed up facilitated by the driving term is related to a sharp increase in the speed of the dynamics near the avoided crossing. In essence, CD seeks to `run' through the difficult points in the evolution, and this leads to an increase in the energy used. This can also been seen by examining the energy spectra of the control Hamiltonians themselves which are shown in Fig.~\ref{fig1}(c) where the solid curves correspond to the ground and excited state for $H\!=\!H_0+H_\text{CD}$. We see that the addition of the control field leads to the evolving Hamiltonian having an increasingly large gap. Thus, as the system is evolved according to this new Hamiltonian, it can be driven progressively faster until it reaches the avoided crossing of the original Hamiltonian. The dashed curves show the corresponding energy eigenvalues for the LCD Hamiltonian, where a similar behavior is observed throughout except at the start and end of the ramp. In contrast, BOB essentially does not have to deal with the difficulties that arise when approaching the avoided crossing as it mostly evolves according to a system with no applied field. It is interesting that when the system is evolving near the avoided crossing the instantaneous cost for both CD and LCD exhibit an identical behavior despite their respective evolved states differing greatly at these times, cf. Fig.~\ref{fig1}{(a)}. We find notable differences in the behavior of the instantaneous costs appear only in the earlier and later stages of the protocols and again these features are reflected in the respective energy eigenvalues of the applied Hamiltonians. These differences are very sensitive to the total protocol duration, as shown in Fig.~\ref{fig1}{(d)}-{(f)} where we we show the same quantities for $\tau=0.1$ (notice that this duration is significantly shorter than the QSL time and therefore OC only varying the field is not possible). For fast driving the LCD has very high instantaneous costs, while for longer protocols we find $\partial_t \mathcal{C}$ can be lower for LCD compared to CD.

\begin{figure}[t]
\includegraphics[width=0.8\columnwidth]{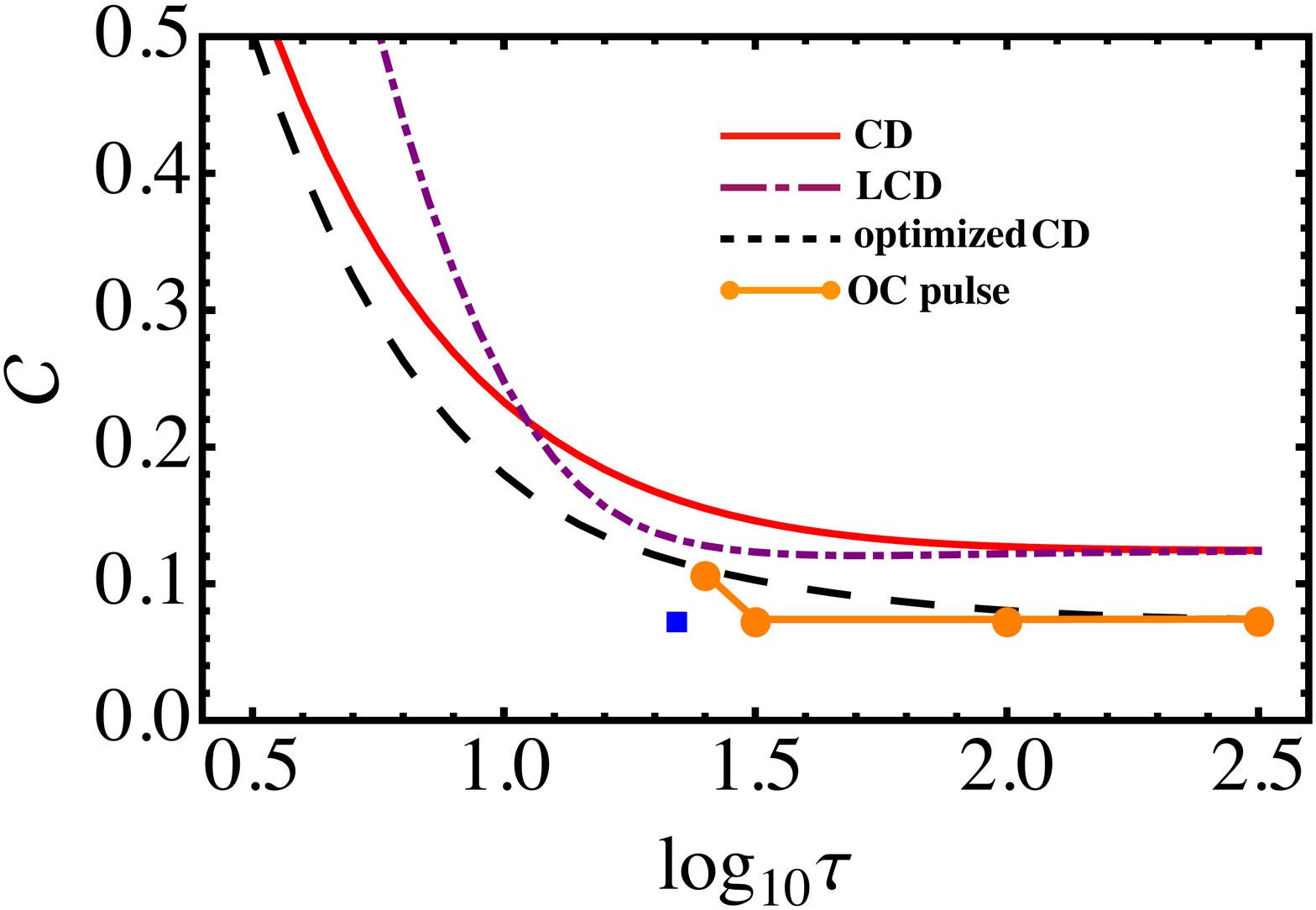}
\caption{Cost versus the protocol duration, $\tau$, for CD (solid, red) and LCD (dot-dashed, purple) when the ramp takes the form given by Eq.~\eqref{LCDramp}. For sufficiently slow protocols, we see a cross over in the over all energy used. The dashed black curve corresponds to the optimized ramp for CD given by Eq.~\eqref{TQDoptimized} taking $\epsilon=0.1$ and $m=40$. Orange circles correspond to the OC pulse of Eq.~\eqref{eq:gocfourier} with an infidelity $q<10^{-9}$. The square point is the cost for the BOB pulse Eq.~\eqref{eq:bangoffbang} resulting in unit fidelity.}
\label{fig3}
\end{figure}

This last observation has an interesting consequence: if we compute the total cost, by integrating Eq.~\eqref{cost}
\begin{equation}
\label{CostInt}
\mathcal{C} =\frac{1}{\tau}\int_0^\tau \| H_k \| dt,
\end{equation}
 we find that for fast protocols using the ramp given by Eq.~\eqref{LCDramp} CD is less costly, as shown in Fig.~\ref{fig3}. However there is a crossover. For sufficiently slow processes, but still faster than the adiabatic limit, LCD becomes the less resource intensive control method. The crossover is dependent on the value of $\Delta$; we find that larger values lead to a crossover at smaller $\tau$.

In Fig.~\ref{fig3}, we also add results obtained with an OC method that is applicable for $\tau\!>\!\tau_\text{QSL}$ in which the time dependence of the control field is the sum of a linear ramp and a truncated Fourier series as
\begin{equation}
\label{eq:gocfourier}
g_\text{OC}(t) = g_0 - 2g_0\frac t\tau + \sum_{n=1}^{n_{\rm max}} 
a_n \sin\left(\frac{n\pi t}{\tau}+\phi_n\right).
\end{equation}
where $n_{\rm max}$ is the maximum number of Fourier components. To obtain the optimized parameters $\{a_n,\phi_n\}$ and thus the function $g_\text{OC}(t)$, we numerically minimize the combined function: $q^\gamma \mathcal C$ where $q\!=\!1-\mathcal F$ is the final infidelity. We tune the power $\gamma$ to best minimize simultaneously the infidelity and the cost $\mathcal C$. In our calculations we choose $10^{-3}\!<\!\gamma\!<\!10^{-2}$ and $20\!<\!n_{\rm max}\!<\!50$ depending on $\tau$. For $\tau\!>\!\tau_\text{QSL}\!\sim\! 22.14$, we are consistently able to achieve a very small infidelity of $q<10^{-9}$. We find that OC is always the minimally resource intensive control technique, and more remarkably, for the considered Landau-Zener model here, the total cost appears almost independent of the protocol duration.

In the limit of $\tau\!\to\!\tau_\text{QSL}$, we find this particular OC method becomes less efficient as the number of frequencies to be retained increases very rapidly, thus indicating that the form of control pulse is becoming progressively harder to realize using a smooth function. Indeed, we know that at $\tau_\text{QSL}$ the required ramp is given by the BOB pulse. Of course, this could in principle be emulated by Eq.~\eqref{eq:gocfourier} for sufficiently large $n_{\rm max}$. We show the total cost for the BOB pulse in Fig.~\ref{fig3} (blue square) and the agreement with the OC results obtained at larger $\tau$ is clearly evident, thus indicating the claimed invariance of the cost of OC to protocol duration. Finally, we remark that OC could also be directly applied to when both $\Delta$ and $g$ vary time-dependently. In such a case it is likely that $\tau\!<\!\tau_\text{QSL}$ is achievable, however, evidently, this is a more involved scenario.

While we have restricted to using the smooth ramp Eq.~\eqref{LCDramp}, as noted previously, CD allows for any ramp to be used. Thus, unlike in typical OC  methods where fidelity is maximized for a given protocol duration, since CD already guarantees perfect fidelity, we are able to optimize the choice of ramp which minimizes the cost. We refer to Ref.~\cite{MortensenNJP} where a similar approach is successfully implemented. The cost, Eq.~\eqref{CostInt} can be expressed in terms of $s\!\!=\!\!t/\tau$ as
\begin{eqnarray}
\mathcal{C}&=&\frac{1}{\tau} \int_{0}^{\tau} \| H(t) \| dt = \int_{0}^{1} \| H_{0}(s)+ \tau^{-1} H_\text{CD}(s) \| ds \nonumber \\
&=& \int_{0}^{1} \left[ \sum_n E_n^2(s)+ \tau^{-2} \sum_{\substack{n,a \\ n\neq a}} |A_{n,a}(s)|^2  \right]^{1/2} ds,
\end{eqnarray}
where $A_{m,n}\!=\!\frac{i \bra{m} \left( \partial_t H_0 \right)\ket{n}}{E_n-E_m}$ and we have used the Frobenius norm as before. Clearly the cost scales with the total time $\tau$ and the contributions from the different Hamiltonian terms $H_0$ and $H_\text{CD}$ is evident. Note that the contribution from $H_\text{CD}$ is similar to the usual criteria for adiabaticity i.e. $\sum_{n \neq m}  \left| A_{m,n}\right|^2  \ll 1$ as one would expect.

In the non-adiabatic regime (i.e. small total time $\tau$)
\begin{eqnarray}
\mathcal{C} \approx \tau^{-1}  \int_{0}^{1} \| H_\text{CD}(s) \| ds.
\end{eqnarray}
If we define $\| H_\text{CD}(s) \|$ as a Lagrangian and minimising the corresponding action, we find the ramp with the lowest cost in this regime to be given by
\begin{equation}
g_{NA}(s)= \Delta \tan \left[ c_1 \Delta (s+c_2)\right], 
\end{equation}
where $c_{1,2}$ are constants of integration which are fixed by the boundary conditions of $g(t)$. The cost in this case is $\mathcal{C} \approx \frac{|c_1 \Delta|}{\sqrt{2}\tau}$. The dependence on the energy gap shows that for large $\Delta$, the ramp tends toward a linear pulse while for small $\Delta$ it tends towards a delta pulses at each endpoint. Similarly, in the adiabatic regime (i.e. large $\tau$) we have,
\begin{eqnarray}
\mathcal{C} &\approx&  \int_{0}^{1} \| H_{0}(s) \| ds \label{adiab_approx} \\
                        &=& \int_{0}^{1}  \sqrt{(\Delta^2+g(s)^2)/2} \; ds.
\end{eqnarray}
Thus the optimal ramp is a delta pulse at the endpoints, in order to fulfil the boundary conditions, and zero otherwise, so the positive integrand is as small as possible. This can be approximated by a continuous function as
\begin{eqnarray}
g_{A}(s)=-g_0 \left[ \tanh(m s- m)+\tanh(m s)\right] \label{adiab_opt}
\end{eqnarray}
for $m\! \gg \!1$. It simply remains to tackle intermediate values of $\tau$. The results for the two regimes can be combined as
\begin{equation}
\label{TQDoptimized}
g_C(s,\tau)=f(\tau) g_{A}(s)+ \left[1-f(\tau)\right] g_{NA}(s)
\end{equation}
for monotonically increasing $f(\tau)$ bounded between $0$ and $1$. One choice is $f(\tau)\!\!=\!\!\frac{2}{\pi} \arctan(\epsilon \tau)$, where $\epsilon$ determines how fast one changes from one regime to another. This approach has the advantage that the total time $\tau$ does not have to be accounted for when determining the optimal pulse. It may also prove useful in cases where calculating the norm of the full Hamiltonian is difficult but estimating the norm of the adiabatic and counterdiabatic components is more tractable.

 In Fig.~\ref{fig3}, the dashed black curve corresponds to CD when the ramp is given by Eq.~\eqref{TQDoptimized}. Thus, for the paradigmatic Landau-Zener model we see that OC is the lowest cost control technique, however, CD and LCD offer some noticeable advantages. Using these techniques, the final state fidelity is guaranteed independent of the protocol duration and these methods are applicable when arbitrarily fast control is required. For CD the associated cost can be further minimized with respect to the form of ramp employed. While this is still more costly than OC, it allows for driving times faster than the QSL which are not achievable using OC techniques that depend solely on manipulating the applied field.

\begin{figure*}
\hskip0.65cm{\bf (a)} \hskip0.65\columnwidth {\bf (b)} \hskip0.65\columnwidth {\bf (c)}\\
\includegraphics[width=0.65\columnwidth]{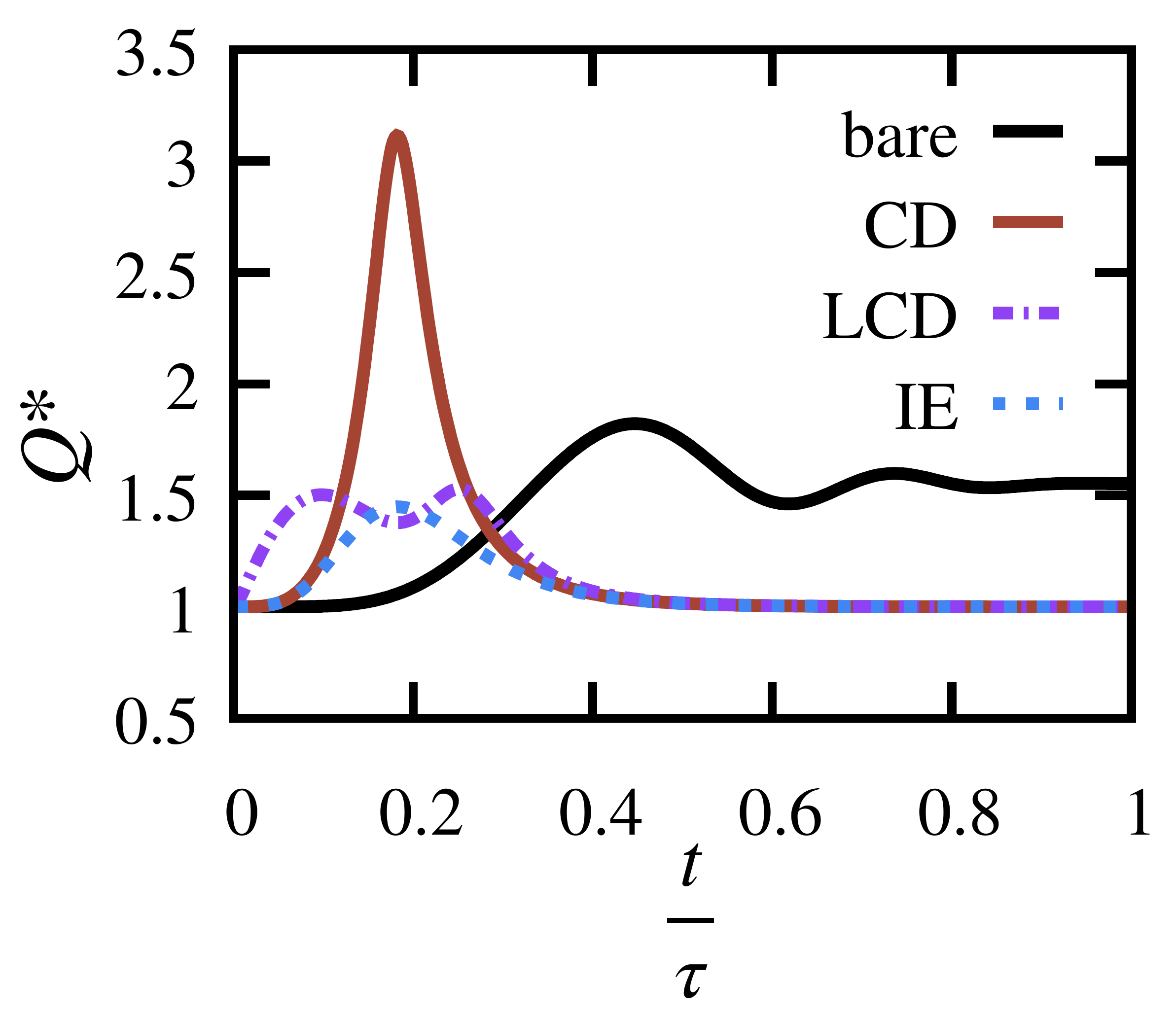}
\includegraphics[width=0.65\columnwidth]{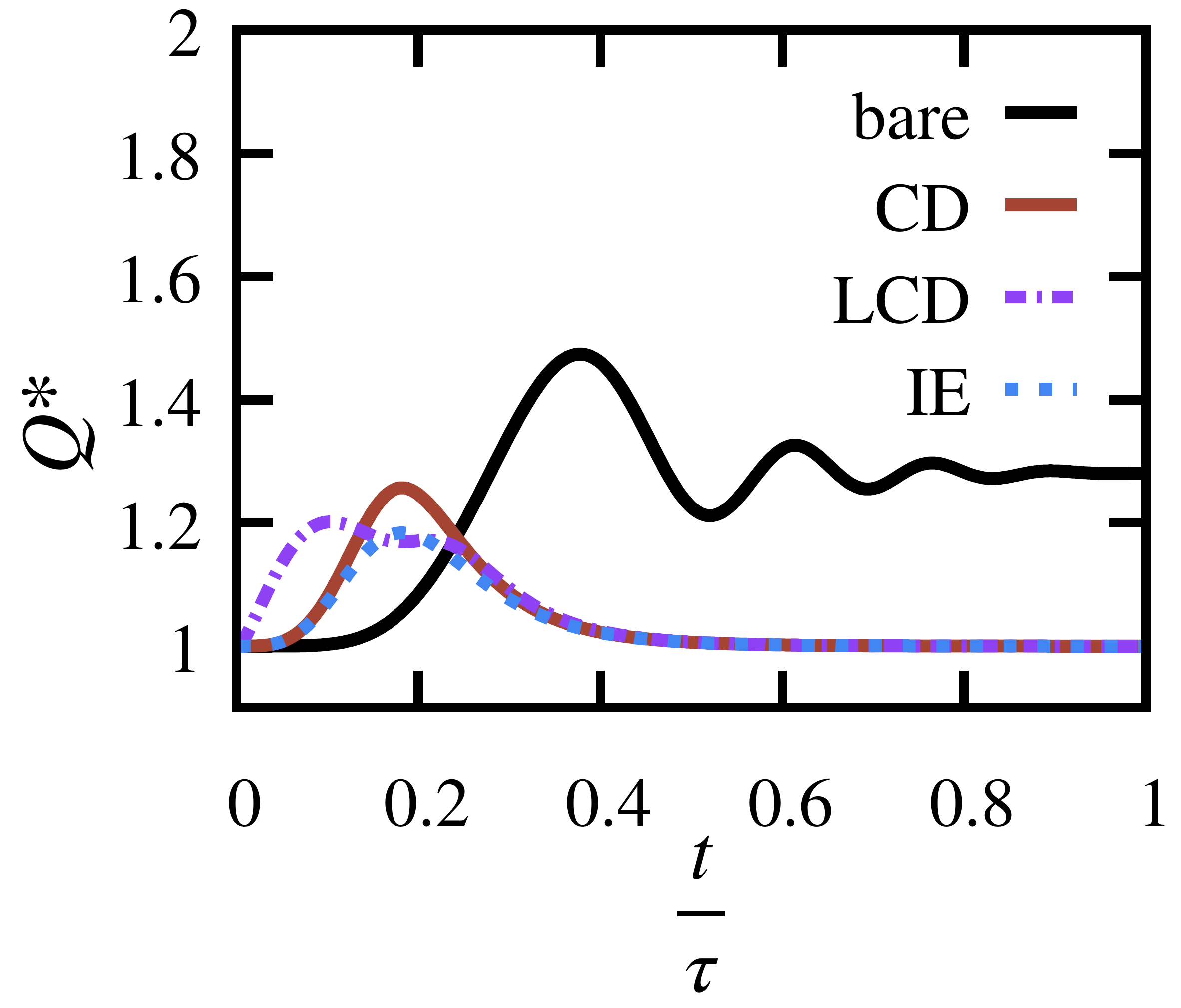}
\includegraphics[width=0.70\columnwidth]{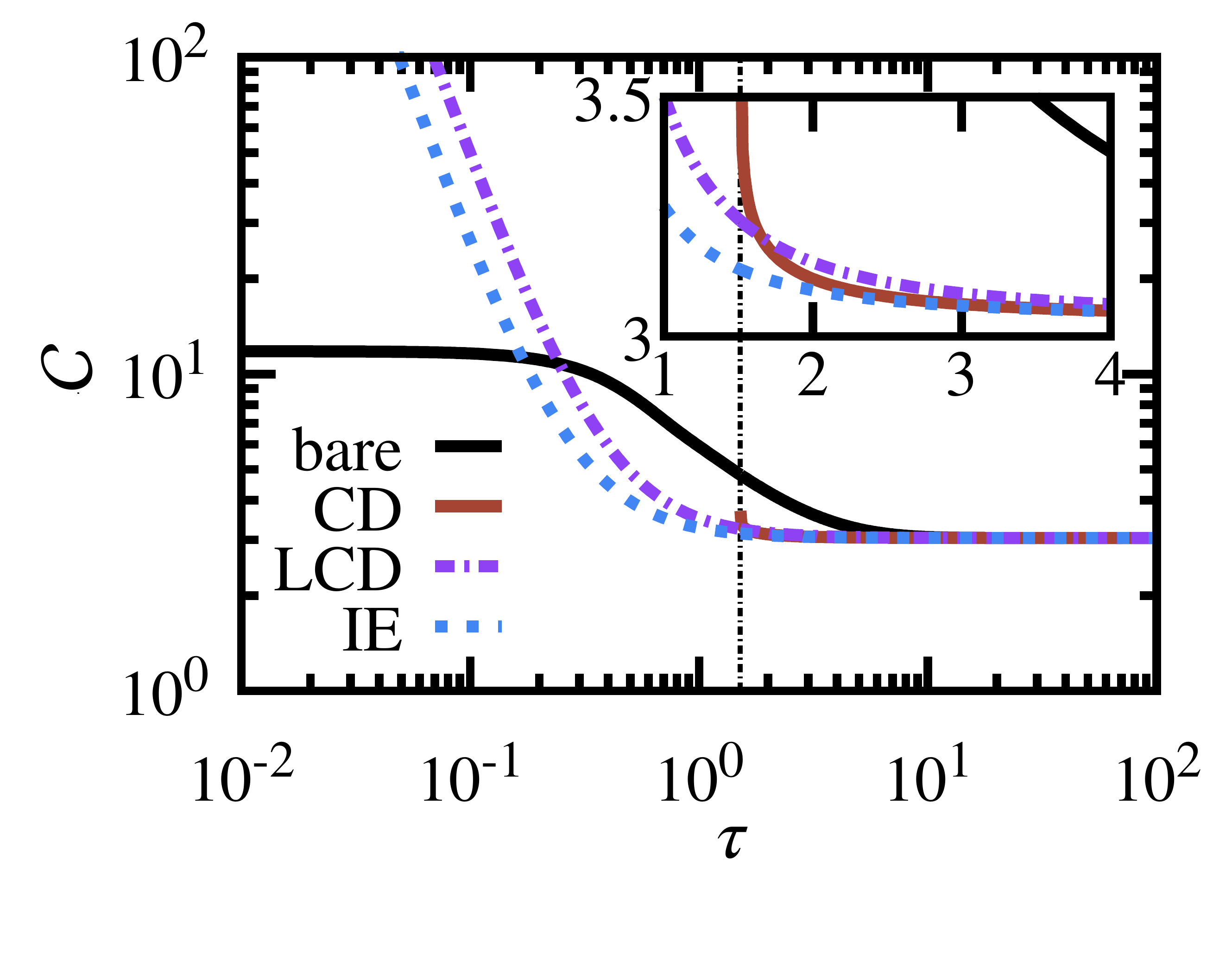}
\caption{Quantum harmonic oscillator. {\bf (a)} Adiabaticity parameter $Q^\ast$ of various control protocols as a function of time, assuming $\omega(t)$ takes the form in Eq.~\eqref{HOramp}. The adiabaticity parameter $Q^\ast$ of bare Hamiltonian $H_0$ (solid black) is compared with LCD (dot-dashed, purple), CD (red) and IE (dotted, blue) for $\tau\!=\!1.6$. {\bf (b)} The adiabaticity parameter $Q^\ast$ for $\tau = 2.5$.  {\bf (c)} Cost for the full CD Hamiltonian, $H_0+H_\text{CD}$, (solid, dark red), the LCD Hamiltonian, $H_\text{LCD}$, (dot-dashed, purple) and the IE Hamiltonian, $H_\text{IE}$, (dotted, blue). The inset shows a zoom when LCD becomes less costly and the vertical dotted line corresponds to the time, $\tau\! \approx\! 1.52$ that constrains the CD protocol to ensure no trap inversion occurs. In all panels we fix the initial frequency $\omega_0 = 1$, final frequency $\omega_1 =10$ and the inverse temperature $\beta =1/(k_B T) = 3$.}
\label{fig4}
\end{figure*}
\subsection{Parametrically Driven Quantum Harmonic Oscillator}
Let us now consider the case of a time-dependent harmonic oscillator, initially in thermal equilibrium at inverse temperature $\beta\!=\!1/(k_B T)$, with mass $m$ whose Hamiltonian is of the usual form,
\begin{equation}
H_0 =\frac{p^2}{2m} + \frac{m}{2} \omega^2(t) x^2,
\end{equation}
where $x$ and $p$ are the position and momentum operators, respectively, and we assume that the time-dependent frequency $\omega(t)$ starts with initial value $\omega_0$ at $t\!=\!0$ and ends with final value $\omega_1$ at $t\! =\! \tau$. The state of the oscillator remains Gaussian for any driving protocol $\omega(t)$ due to the quadratic form of the Hamiltonian. The Schr\"odinger equation for the parametric quantum harmonic oscillator can be solved exactly for any frequency modulation~\cite{Husimi1953,Deffner2008PRE, Deffner2010CP}. The system dynamics are completely determined by a dimensionless adiabaticity parameter, $Q^\ast$, as introduced by Husimi \cite{Husimi1953},
\begin{equation}
Q^\ast = \frac{1}{2\omega_0 \omega(t)}\left\{ \omega^2_0 \left[\omega^2(t)\,X_\tau^2 + \dot{X}^2_\tau\right] + \left[\omega^2(t)\,Y_\tau^2 + \dot{Y}_\tau^2\right] \right\},
\label{Qast}
\end{equation}
where $X_t$ and $Y_t$ are the solutions of the force-free classical oscillator equation, $\ddot{X}_t + \omega^2(t) X_t \!=\!0$, satisfying the boundary conditions $X_0 \!=\!0$, $\dot{X}_0\!=\! 1$ and $Y_0\!=\! 1$, $\dot{Y}_0 = 0$. The adiabaticity quantity $Q^\ast \ge 1$ is the ratio of the nonadiabatic mean energy and the adiabatic energy and is equal to one for slow driving that realizes adiabatic transformations. 

Using Eq.~\eqref{berryeq} we can determine the CD term~\cite{Muga2010JPB}
\begin{equation}
H_\text{CD} = -\frac{\dot{\omega}(t)}{4\omega(t)} (xp + px),
\end{equation}
and consequently the adiabaticity parameter can be expressed as~\cite{Mishima2017}
\begin{equation}
Q^\ast_\text{CD} =\left[1 - \frac{\dot{\omega}^2(t)}{4 \omega^4(t)}\right]^{-1/2}.
\end{equation}
We note that the time variation of the frequency must fulfil the condition, $\omega^2(t) \! > \! \dot{\omega}^2(t)/[4 \omega^2(t)]$, to avoid the trap inversion. This is consistent with the conditions in typical experimental realizations and also ensures that the adiabaticity criterion Eq.~\eqref{Qast} retains a clear physical interpretation during the process.

Considering the LCD approach \cite{STAreview, delCampo2013PRL, Deffner2014PRX}, similarly to the qubit case, the nonlocal CD term is mapped onto a unitarily equivalent Hamiltonian with a local potential of the form 
\begin{equation}
H_\text{LCD} = \frac{p^2}{2m} + \frac{m\Omega^2(t) x^2}{2},
\label{19}
\end{equation}
with the modified time-dependent squared frequency,
\begin{equation}
{\Omega}^2(t) = \omega^2(t) -\frac{3\dot{\omega}^2(t)}{4 \omega^2(t)}+\frac{\ddot{\omega}(t)}{2 \omega(t)}. 
\label{20}
\end{equation}
The exact dynamics of the system are obtained from the solution of the adiabaticity parameter, Eq.~\eqref{Qast} solved by replacing $\omega(t)$ with $\Omega(t)$. Again, to avoid the inversion of the harmonic trapping potential, the effective frequency $\Omega(t)$ must be positive ($\Omega^2(t) > 0$).

A final control technique that is particularly effective for the oscillator case is inverse engineering (IE) based on constructing appropriate parameter trajectories of the frequency by employing the Lewis-Riesenfeld invariants of motion~\cite{Lewis1969}. Considering $H_0$, the dynamics are obtained by solving the Schr\"odinger equation based on the invariants of motion of the following form \cite{Chen2010PRL,Chen2010PRA},
\begin{equation}
I(t) = \frac{1}{2}\left(\frac{{x}^2}{ b^2}m \omega_0^2 + \frac{1}{m}{\pi}^2\right),
\end{equation}
where ${\pi} =  b{p} - m\dot{b} {x}$ plays the role of a momentum conjugate to ${x}/b$, $\omega_0$ and is, in principle, an arbitrary constant taken as $\omega_0 \!=\! \omega(0)$, and the dimensionless scaling function $b(t)$ satisfies the Ermakov equation
\begin{equation}
\ddot {b}(t) + \omega^2(t)  b(t) = \omega_0^2/b^3(t).
\label{23}
\end{equation}
The resulting time-dependent instantaneous energy of the Hamiltonian reads
\begin{equation}
\langle H_\mathrm{IE} (t)\rangle = \frac{1}{2} \left[\frac{\dot{b}^2(t)}{2\omega_0} + \frac{\omega^2(t)  b^2(t)}{2 \omega_0} + \frac{\omega_0}{2 b^2(t)}\right] \coth\left(\frac{\beta  \omega_0}{2}\right),
\end{equation}
and corresponding adiabaticity parameter is given by \cite{Abah2018PRE}
\begin{equation}
Q_\mathrm{IE}^{\ast}(t) = 1 + \frac{\dot{\omega}^2(t)}{8 \omega^4(t)}. 
\end{equation}

The behavior of the various adiabaticity parameters for the three considered control protocols is shown in Fig.~\ref{fig4} using a ramp analogous to Eq.~\eqref{LCDramp}~\cite{STAreview}
\begin{equation}
\label{HOramp}
\omega(t) = \omega_0 + 10 \omega_d \left(\frac{t}{\tau}\right)^3 - 15 \omega_d \left(\frac{t}{\tau}\right)^4 + 6 \omega_d \left(\frac{t}{\tau}\right)^5,
\end{equation}
where the difference between final and initial frequency is $\omega_d = \omega_1 - \omega_0$. We clearly see a similarity between the methods as they all start and end at the same value of adiabaticity. However, the CD has the largest fluctuation while the behaviour of the LCD and IE show significantly smaller peaks. The IE technique gives the smallest value of adiabaticity parameter which results in the smallest nonadiabatic excitation during the process. In Fig.~\ref{fig4}(a) we show the adiabaticity parameter for a shorter time duration ($\tau = 1.6$) and observe a large increase in the nonadiabatic excitations than the case of $\tau = 2.5$, see Fig.~\ref{fig4}(b). The validity CD protocol, as dictated by the constraint ensuring no trap inversion occurs, breaks down at time $\tau\! \approx\! 1.52$.

\begin{figure*}[!t]
\hskip2cm{\bf (a)} \hskip0.55\columnwidth {\bf (b)} \hskip0.7\columnwidth {\bf (c)}\\
\includegraphics[width=1.3\columnwidth]{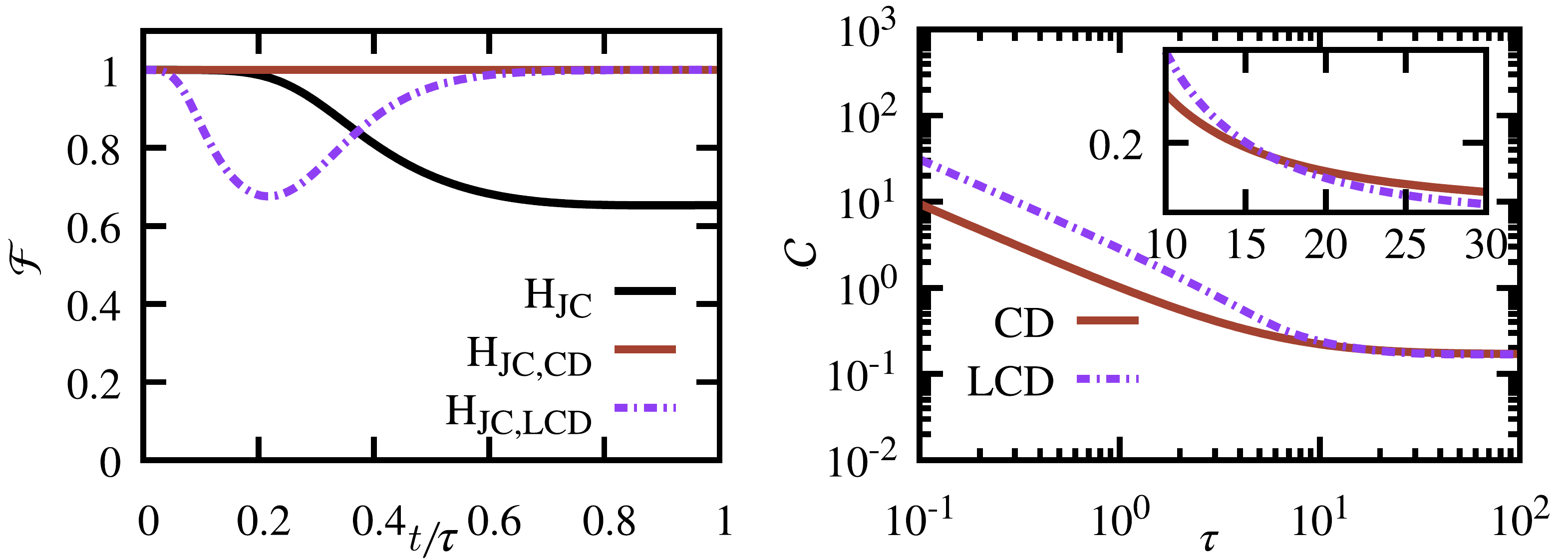}\includegraphics[width=0.63\columnwidth]{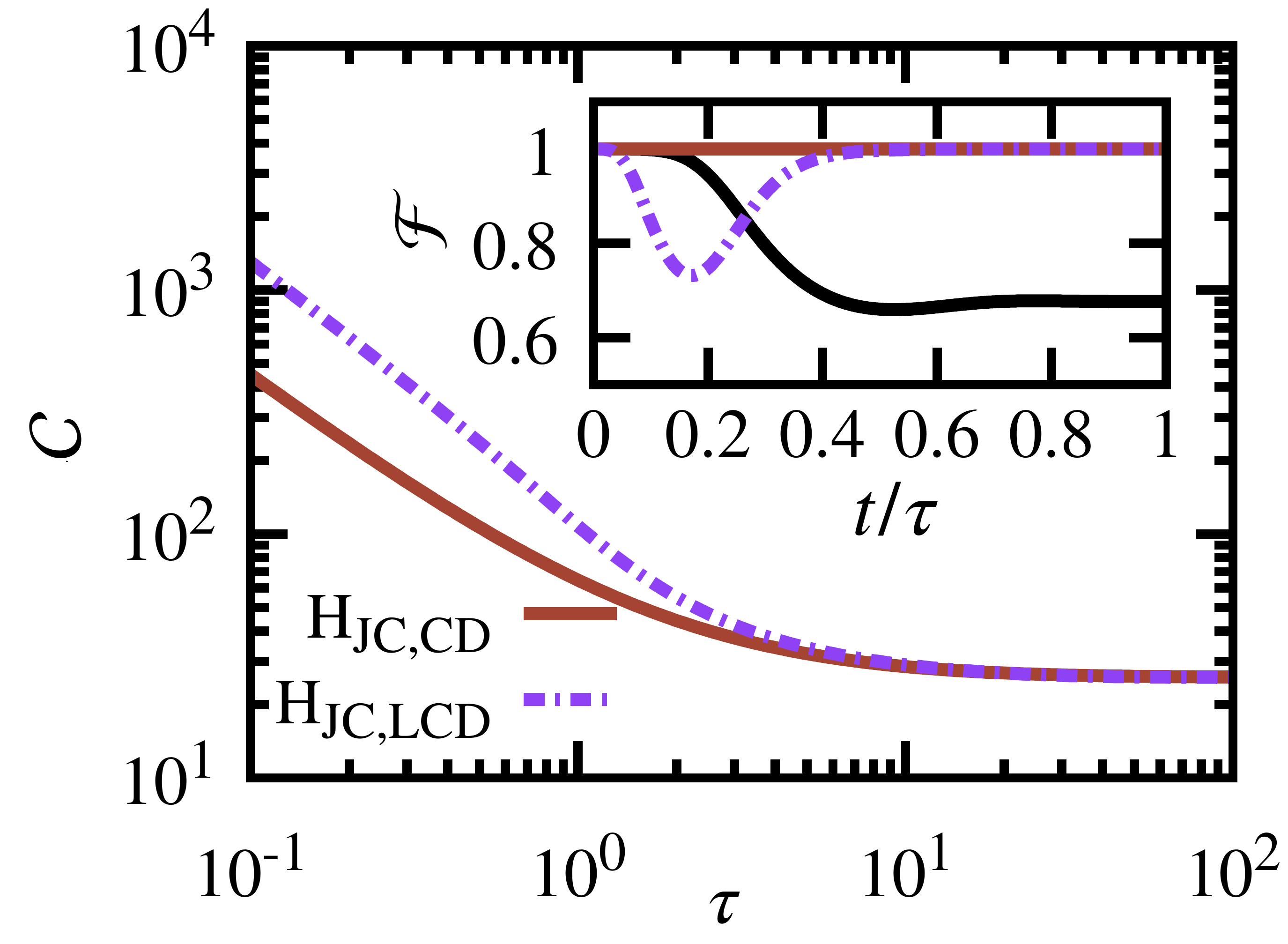}
\caption{Examples for the Jaynes-Cummings model. {\bf (a)} Fidelity ${\cal F}(t)$ with respect to the instantaneous ground state of the corresponding original Hamiltonian, assuming $g(t)$ of the form in Eq.~\eqref{LCDramp}. Bare Hamiltonian evolution (solid black curve) is compared with CD (solid dark red curve) and LCD (dot-dashed purple curve) protocols. Here, the total time of the protocol is $\tau=10$. {\bf (b)} Cost $\mathcal{C}_k$ as a function of the total time $\tau$ (in units of $\omega$) for the ramp in Eq.~\eqref{LCDramp}. The inset shows a zoom when LCD becomes less costly than the CD (at around $\tau\approx 17$). {\bf (c)} Jaynes-Cummings model with an initial coherent state $\ket{\alpha}$ of the field (the total initial state of the system is thus $\ket{e,\alpha}$). Cost $\mathcal{C}_k$ as a function of the total time $\tau$ (in units of $\omega$) for the ramp in Eq.~\eqref{LCDramp} and $\alpha = 2$. The inset shows the fidelity with respect to the instantaneous ground state of the corresponding Hamiltonian for the total time $\tau = 10$ with the same color conventions as the previous plots. The parameters used are $g(0)=0$, $g(\tau)=0.2\omega$, $\delta=0.1\omega$ and $\omega=1$.}
\label{fig5}
\end{figure*}
Turning our attention to the instantaneous cost as defined in Eq.~\eqref{cost}, as the spectrum is unbounded, the resulting norm is not finite. To circumvent this issue we note that the the average energy of the system evaluated over the the full Hamiltonian will behave in a qualitatively identical manner and as such we use it as an indicator of the control cost. Thus for the oscillator, we must modify our definition of the total cost to
\begin{equation}
\mathcal{C} = \frac{1}{\tau} \int_0^\tau \langle H_\mathrm{tot}\rangle dt,
\end{equation} 
where the full Hamiltonian for a given control protocol is~\cite{Abah2018PRE}
\begin{equation}
\langle H_\mathrm{tot}\rangle = \frac{\omega(t)}{\omega_0} Q_k^\ast \langle H(0)\rangle = \frac{ \omega_t}{2} Q_k^\ast \coth\left(\beta  \omega_0/2\right)
\end{equation}
with $k \!=\! \mathrm{CD, LCD, IE}$. We note, for the LCD, the $\omega(t)$ is replaced with $\Omega(t)$ in the equation above as well as in evaluation of $Q^\ast$ with Eq.~\eqref{Qast}. In Fig.~\ref{fig4}(c) we numerically evaluate the cost of the evolution for the various control protocols using ramp Eq.~\eqref{HOramp}. We observe that while all the protocols lead to the same value of cost for long durations, they significantly differ for fast processes. We find that IE is the most efficient of the three protocols. Furthermore, in line with the Landau-Zener model, for intermediate timescales the CD performs better than the LCD but there is a crossover as the driving time becomes smaller. Thus, our results indicate that a qualitatively similar hierarchy emerges in the case of driving a thermal harmonic oscillator.

\subsection{Jaynes-Cummings Model}\label{s:JCM}
As a final case study, we examine the Jaynes-Cummings model~\cite{Jaynes1963IEEE}. Recently, owing to its richness, this model has attracted renewed interest in diverse areas such as quantum control~\cite{Gerry2005Book,Barnett2017PQE}. We thus consider the model
~\cite{Joshi1993PRA,Lawande1994PRA}
\begin{equation}
H_{JC} =  \frac{\omega_A}{2} \sigma_z +   \omega a^\dagger a +   g(t) (a\sigma^+ + a^\dagger \sigma^-),
\label{jcmodel}
\end{equation}
which describes the interaction of a two-level atom, modelled as a spin-$\frac{1}{2}$ particle, with a single mode of the electromagnetic field whose annihilation and the creation operators are $a$ and $a^\dagger$, respectively. While the free Hamiltonian $H_0\!=\!  \omega_A \sigma_z/2 +    \omega a^\dagger a$ is assumed to be time-independent, the interaction Hamiltonian $H_{int}\!=\!  g(t) (a\sigma^+ + a^\dagger \sigma^-)$ depends on a  time-dependent coupling rate $g(t)$, upon which we exert control~\cite{Joshi1993PRA,Lawande1994PRA,Law1995PRA}. As the total number of excitations in the system ${N}_e\!\equiv \!\ket{e}\bra{e} + a^\dagger a$ is a constant of motion, for any given initial number of photons $n$ in the field the dynamics is restricted to the subspace spanned by states $\{\ket{e,n}, \ket{g,n+1}\}$. Owing to this feature Eq.~\eqref{jcmodel} may be written as the direct sum of Hamiltonian terms $(H_{JC})_n$ labelled by the corresponding number of photons in the field. Over the basis $\{\ket{e,n}, \ket{g,n+1}\}$, such terms take the form
~\cite{Shore1993JMO,Gerry2005Book}
\begin{eqnarray}
(H_{JC})_n
&=& \frac{1}{2} \left(\begin{array}{cc} (2n+1)\omega + \delta & 2 g(t) \sqrt{n+1}\\ 2 g(t) \sqrt{n+1}  &(2n+1)\omega - \delta \end{array} \right), 
\label{HJCn}
\end{eqnarray}
where  $\Omega_{R}(t)\equiv2g(t) \sqrt{n+1}$ is the $n$-photon time-varying Rabi frequency and $\delta\! =\! \omega_A - \omega$ corresponds to the detuning parameter of the radiation from the atomic resonance.

Eq.~(\ref{HJCn}), can be mapped to a Landau-Zener model by introducing the spin-like operators $\bar{\sigma}^-\!=\!\ket{g,n+1}\bra{e,n}$, $\bar{\sigma}^+\!=\!\ket{e,n}\bra{g,n+1}$, $\bar{\sigma}_z\!=\!\ket{e,n}\bra{e,n}-\ket{g,n+1}\bra{g,n+1}$, so that
\begin{align}
(H_{JC})_{n}=\frac{(2n+1) \omega}{2}\mathbb{I}+\frac{\delta}{2}\bar{\sigma}_z+  \frac{ \Omega_R(t)}{2}\bar{\sigma}_x.
  \end{align}
Moreover, through a $\pi/2$ rotation about the $y$-axis, we have $\bar{\sigma}_z\rightarrow \bar{\sigma}_x$ and $\bar{\sigma}_x\rightarrow -\bar{\sigma}_z$, which takes us to 
\begin{align}\label{eq:HnLZ}
  (H_{JC})_{n}=\frac{(2n+1) \omega}{2}\mathbb{I}+\frac{\delta}{2}\bar{\sigma}_x-\frac{ \Omega_R(t)}{2}\bar{\sigma}_z.
  \end{align}
From Eq.~(\ref{eq:HnLZ}), we can already see that Eqs.~(\ref{qubitTQD}) and~(\ref{qubitLCD}) are valid upon the identification of $\Delta\rightarrow \delta$ and $g(t)\rightarrow -\Omega_R(t)$.

The desired CD Hamiltonian corresponding to this problem is given by~\cite{Demirplak2003,Berry2009JPA}
\begin{equation}
\label{JCcd}
\begin{aligned} 
H_\text{CD} &= i \sum_{n,\sigma=\pm}  (\partial_t\ket{ (n,\sigma(t))} \bra{n,\sigma (t)} \\ &-  \bra{n,\sigma (t)}\partial_t \ket{(n,\sigma(t))} \ket{n,\sigma (t)}\bra{n,\sigma (t)}), \nonumber\\
&=\dot{\theta}_n(t) \bar{\sigma}_y
\end{aligned}
\end{equation}
with the mixing angle $\theta_n(t)\!=\! \frac{1}{2} \arctan \left(\frac{\Omega_{R}(t)}{\delta}\right)$ and $\ket{n,\sigma (t)}$ denoting the dressed-atom eigenstates of the original Hamiltonian. The explicit expressions of the new modified total Hamiltonian for CD and LCD are presented in the Appendix.

Considering again the smooth ramp of $g(t)$ in the form of Eq.~\eqref{LCDramp} and for the initial state initially $\ket{e,0}$, a unitary evolution is performed from $g(0) \!=\! 0 $ to a target state at $g(\tau)\!=\!0.2\omega$ with initially fixed $\delta\!=\!0.1\omega$ and setting $\omega\!=\!1$. In Fig.~\ref{fig5}(a) we show the fidelity of the state evolving according to CD, LCD, and the bare Hamiltonian using ramp Eq.~\eqref{LCDramp}, with respect to the instantaneous ground state of $(H_{JC})_{n=0}$ for $\tau\!=\!10$. Again, we find a qualitative similarity in the behaviour of CD and LCD when compared with the Landau-Zener model.

In Fig.~\ref{fig5}(b) we show the cost by applying Eqs.~\eqref{cost} and~\eqref{CostInt} to such model
for both a CD and LCD strategy and $n=0$ excitations (we neglect constant energy factors of $H_{JC,{\rm CD}}$ and $H_{JC,{\rm LCD}}$),
finding it again qualitatively in line with was observed for the Landau-Zener model [cf.~Fig.~\ref{fig3}].

In Fig.~\ref{fig5}(c) we examine the cost and fidelity of the state evolving according to CD and LCD strategies starting from a coherent state $\ket{\alpha}$ of the cavity field with the ramp in Eq.~\eqref{LCDramp} and the amplitude $\alpha=2$. Clearly we see a similarity in the behaviour of both shortcut protocols with the case of vacuum initial state in perfectly achieving the target state [cf. inset of Fig.~\ref{fig5}(c)]. However, the cost of shortcut to adiabaticity protocols are higher than the vacuum state situation as more $n$-subspaces must be considered, in light of the form of the initial state of the field. For our calculations we have computed the cost and the fidelity using $n=0,\dots,40$. Such a cutoff is well justified as the populations of the states $\ket{g,m}$ and $\ket{e,m}$ with $m\!>\!40$ are $p_{m>40}\!<\! 10^{-20}$. In keeping with the previous results we once again find that, for shorter protocol durations, CD is energetically more efficient than LCD, while for larger values of $\tau$, the LCD strategy becomes less costly. 

\section{Conclusions}
\label{conclusion}
We have quantitatively compared and contrasted the energetic cost of achieving finite time adiabatic dynamics in a variety of physically relevant settings, namely the Landau-Zener model, the parametric quantum harmonic oscillator, and the Jaynes-Cummings model. By exploiting a cost function based on the norm of the driving Hamiltonian~\cite{Zheng2016PRA}, we have shown that a hierarchy in the resource intensiveness emerges. For the Landau-Zener model, we have shown that optimal control protocols appear to be the most efficient techniques and presented a remarkable invariance to the protocol duration. Conversely, counter-diabatic driving was shown to be more costly, however it allows for arbitrarily fast manipulation. We showed the manipulation of a system beyond the quantum speed limit is possible only when the system energy spectrum is affected, precisely as is the case for local and full counterdiabatic drivings. We found that the general features exhibited in the Landau-Zener case are also present in other physically relevant settings. While we have focused on one particular definition of cost, we nevertheless expect our results to qualitatively hold for other suitable choices, such as those based on excess energy~\cite{Abah2017EPL} or work fluctuations~\cite{Funo2017PRL}. Our analysis sheds light on the relative effectiveness of promising strategies for the control of quantum dynamics. By highlighting the respective advantages of such strategies, and the associated cost, the information provided by our study will be useful in conjunction with complementary studies on the achievable minimal control time of quantum dynamics~\cite{poggiPRA2019} for the development of future energy-efficient quantum devices.

\acknowledgements
The authors thank Pablo Poggi and Bar{\i}\c{s} \c{C}akmak for useful discussions. We acknowledge support from the Royal Commission for the Exhibition of 1851, the Basque Country Government (Grant No. IT986-16), the EU Collaborative project TEQ (grant agreement 766900), the DfE-SFI Investigator Programme (grant 15/IA/2864), COST Action CA15220, the Royal Society Newton Fellowship (Grant Number NF160966), the Royal Society Wolfson Research Fellowship ((RSWF\textbackslash R3\textbackslash183013), the Leverhulme Trust Research Project Grant (grant nr.~RGP-2018-266), and the SFI Starting Investigator Research Grant (project ``SpeedDemon", grant nr. 18/SIRG/5508).

\section*{Appendix}
 As discussed in Sec.~\ref{s:JCM}, the Jaynes-Cummings model can be expressed as a direct sum of $2\times 2$-matrix Hamiltonians $(H_{JC})_n$ with $n$ excitations by resorting to the constant of motion $\hat{N}_e$. This allows for a direct identification with a Landau-Zener problem.  For each block, one can construct a CD Hamiltonian, which reads as
\begin{align}
(H_{JC,CD})_n &= (H_{JC})_n + H_{CD}\\&=\frac{(2n+1) \omega}{2}\mathbb{I}+\frac{\delta}{2}\bar{\sigma}_x -  g(t)\sqrt{n+1}\bar{\sigma}_z \nonumber\\&+ \frac{\dot{g}(t)\sqrt{n+1}\delta}{\delta^2+4(n+1)g^2(t)}\bar{\sigma}_y, \nonumber
\end{align}
while the LCD is analogous to Eq.~(\ref{qubitLCD}), which becomes (neglecting a constant energy shift)
\begin{align}
&  (H_{JC,LCD})_n=\frac{1}{2}\sqrt{\delta^2+\frac{4(n+1)\dot{g}^2(t)\delta^2}{(\delta^2+4(n+1)g^2(t))^2}}\bar{\sigma}_x\\
  &-\sqrt{n+1}\left(g(t)+\frac{ (\delta^2+4(n+1)g^2(t))\ddot{g}(t)-8(n+1)g(t)\dot{g}^2(t) }{(\delta^2+4(n+1)g^2(t))^2+4(n+1)\dot{g}^2(t)} \right).
\end{align}
Note that the operators $\bar{\sigma}_{x,y,z}$ refer here to the dressed atom-field basis, namely, $\bar{\sigma}_{x}=\ket{e,n}\bra{e,n}-\ket{g,n+1}\bra{g,n+1}$, $\bar{\sigma}_y=-i\ket{e,n}\bra{g,n+1}+i\ket{g,n+1}\bra{e,n}$ and $\bar{\sigma}_{z}=-\ket{e,n}\bra{g,n+1}-\ket{g,n+1}\bra{g,n+1}$ (see Sec.~\ref{s:JCM}).

\bibliography{hierarchy}

\end{document}